\documentclass{aa}
\usepackage{graphicx}
\def\ssim{\setbox0=\hbox{$\propto$}%
\setbox1=\hbox{$<$}\dimen0=\ht1%
\advance\dimen0by-1.2pt\,\lower.6\dimen0%
\copy0\kern-\wd0\raise.4\dimen0\copy1 \,}

\def\gsim{\setbox0=\hbox{$\propto$}%
\setbox1=\hbox{$>$}\dimen0=\ht1%
\advance\dimen0by-1.2pt\,\lower.6\dimen0%
\copy0\kern-\wd0\raise.4\dimen0\copy1\,}

\def\lambdab{\lambda\mkern-9mu\lower1.2pt\hbox{$\mathchar'26$}}%

\begin{document}
   \title{Diffusion in stellar interiors: critical tests of three
numerical methods}

 \author{Georges Meynet$^1$, Andr\'e Maeder$^1$ and Nami Mowlavi$^{1,2}$}

     \institute{$^1$Geneva Observatory, CH--1290 Sauverny, Switzerland\\
                $^2$Integral Science Data Center, Ch. d'Ecogia, CH--1290 Versoix, Switzerland\\
              email: Georges.Meynet@obs.unige.ch \\
              email: Andre.Maeder@obs.unige.ch\\
              email: Nami.Mowlavi@obs.unige.ch \\     }

   \date{Received ...; accepted ...}

\abstract{ We describe and discuss the properties of three numerical methods for solving the diffusion equation
for the transport of the chemical species and of the angular momentum
in stellar interiors. We study
through numerical experiments both their accuracy and their ability to provide
physical solutions. On the basis of new tests and analyses applied to the stellar
astrophysical context, we show 
that the most robust method to follow the secular evolution
is the implicit finite differences method.
The importance of correctly estimating the diffusion coefficient
between mesh points is emphasized and a procedure for estimating 
the average diffusion
coefficient between a convective and
a radiative zone is described.
}
   
\authorrunning{G. Meynet et al.}
\titlerunning{numerical methods for diffusion}
   \maketitle
%

\section{Introduction}

For many physical processes in stellar interiors 
as heat transfer, transport of angular momentum, mixing
of chemical elements through time dependent convection, semiconvection or
turbulence, a diffusion equation needs
to be solved. Diffusion is thus involved in numerous interesting astrophysical
problems.
For instance,
Vauclair and Charbonnel (\cite{Vau98}) discuss the importance of selective diffusion
in low--metallicity stars and the consequences for lithium primordial abundance; Charbonnel (\cite{Cha95})
has shown how rotational diffusion in low mass stars may lead to the destruction of 
$^{3}$He and thus to new insights on the evolution of the abundance of this cosmological element;
most of the results on the effects of rotation in stellar interiors 
are based on the resolution of a diffusion equation
for both the transport of the chemical species and of the angular momentum
(Chaboyer \& Zahn \cite{Cha92}; Zahn \cite{Za92}; Talon \& Zahn \cite{Ta97}; Denissenkov et al. \cite{Deni99}; 
Heger et al. \cite{He20};Maeder \& Meynet \cite{MM01}; Meynet \& Maeder \cite{MM00}~).
Diffusion plays also a key role in our understanding of 
the structure and the cooling of white dwarfs (cf. Kawaler \cite{Ka95}).

For solving numerically the diffusion equation, different methods are available. Quite
generally, they
can be classed into two main categories: 
the {\it finite differences} methods and the {\it finite elements} methods. Depending on how
the time discretisation is performed, one distinguishes three subclasses in each of these
two main categories, namely the subclasses of the {\it explicit}, {\it implicit} and of the {\it ``Crank--Nicholson''} type methods.
Thus, at least six different methods are available. Here we are interested in modelling the long
term evolution of stars (secular evolution). This immediately prevents us of using
explicit methods which requires the use of much too  short time steps (see below). Among the four
remaining methods, the ``Crank--Nicholson'' finite elements method
was not tested in view of the results obtained by the ``Crank--Nicholson'' finite differences method
(see Sect.~5.4). Thus we shall focus our attention on three of them namely

\begin{itemize}
\item the implicit finite elements method,

\item the ``Crank--Nicholson'' finite differences method,

\item the implicit finite differences method.
\end{itemize}

Descriptions of these methods as well as a general discussion of their
respective advantages and weaknesses 
can be found for instance in Press et al. (\cite{Teu92}) for
the finite differences methods and in Zienkiewicz \& Taylor (\cite{zi00}) for the finite elements methods. 
If these considerations
can help in choosing the best method in general, we think that
very valuable complementary information can be gained 
by applying the methods to the resolution of realistic astrophysical cases,
where the different timescales involved are very specific.
This is what we propose here and this is the main aim of this work.
Let us recall that, as said in Press et al (\cite{Teu92}) ``{\it ... differencing partial
differential equations is an art as much as a science''}, and that
a numerical scheme, which from a theoretical point of view does appear
as very robust can show some weaknesses when applied to a real particular situation. 
This is precisely why it is necessary to test realistic astrophysical situations. This allows us
to make a new and appropriate analysis to support the choice of the best method to handle diffusion problems.

The interested reader will also find in this paper a detailed description of the
discretized equations for resolving the diffusion of chemical elements and of the angular momentum in stars.
We also propose a new way to estimate the diffusion coefficient in the interface between a convective and
a radiative zone.

In Sect. 2 we briefly recall the physical problem to be resolved. The three numerical methods
are described in details in Sect.~3. The estimate of the diffusion coefficient at the interface
between a convective and a radiative zone is presented in Sect.~4. Sect.~5 
is devoted to various tests and comparisons. 
Section 6 summarizes the main results.

\section{The physical problem}

Diffusion is a process by which components of a mixture
move from one part of a system to another as a result of
random motion. Let us briefly recall how the diffusion
velocity is defined (see e.g. Battaner \cite{ba96}).
In a multicomponent fluid, one defines
the mean velocity $\vec v_0$ of the mixture as 

$$\vec v_0={\sum_{i} m_i n_i <\vec v_i>\over \sum_{i} m_i n_i},$$

\noindent where 

$$<v_i> ={1 \over n_i} \int_p f_i \vec v_i {\rm d}\tau_p,$$

\noindent is the mean velocity of the $i$th component, $n_i$
is the partial number density of particles $i$, $m_i$ their mass, $v_i$
their velocity and $f_i(\vec x, \vec p, t)$ the probability distribution function,
which gives the probability to find a particle $i$ at the position $\vec x$,
with the momentum $\vec p$, at the time $t$; $\tau_p$ is the 
phase space volume for the particle's momentum.
One can wonder, why in the expression for $\vec v_0$, the weighting
factor is the partial mass density and not the partial number density.
The question could be formulated in another way:
when is the mean velocity of the mixture zero~? When the net flux
of the number of particles is zero 
($\sum_{i} n_i <\vec v_i>=0$), or when
the net flux of the mass is zero ($\sum_{i} m_i n_i <\vec v_i>=0$)~? Obviously
when the net flux of the mass is zero, otherwise
the centre of gravity of the system is moving and thus the mean velocity of the mixture
is not zero.

The peculiar velocity of a particle is defined as

$$\vec V_i= \vec v_i-\vec v_0.$$

\noindent The mean value of $\vec V_i$ is

\begin{equation}
<\vec V_i>={1 \over n_i}\int_p f_i \vec V_i {\rm d}\tau_p=<\vec v_i>-\vec v_0.
\label{vdif}
\end{equation}

\noindent This quantity is called the diffusion velocity of the $i$th component. It can be
shown very easily, using the expression for $\vec v_0$ that the
different diffusion velocities must compensate one another, {\it i.e.}

\begin{equation}
\sum_i m_i n_i <\vec V_i>=\vec 0.
\label{vcomp}
\end{equation}

\noindent There is thus no net mass flux
associated to diffusion. In that respect, in a star, the changes of chemical composition
due to diffusion can be treated very similarly as those due to the
nuclear reactions. Like the nuclear reactions, diffusion 
modifies the mean molecular weight (and thus
the hydrostatic structure of the star), but does not induce any net mass flux.

In the context
of the Boltzmann microscopic theory, it is
possible to deduce from 
the equations
of motion for the different components, expressions for the
$<\vec V_i>$'s (see Battaner \cite{ba96}). We shall not repeat
these developments here. Instead we consider that
diffusion is equivalent to a displacement of particles whose velocities
may be related to the diffusive coefficient, $D$, by the expression
(see also Sect. 3.2)

$$<\vec V_i>=-{D \over X_i}\vec \nabla X_i,$$

\noindent where $X_i$ is the mass fraction of element $i$. 
The minus sign results from the fact that the velocity
is directed in the opposite direction to the mass fraction gradient. 
This expression automatically satisfies the condition seen above that
$\sum_i m_i n_i <\vec V_i>=\vec 0$, indeed

$$\sum_i m_i n_i <\vec V_i>=-\rho D \vec \nabla \left( \sum_i X_i\right)=\vec 0,$$

\noindent since $\sum_i X_i=1$ and the relation between $n_i$
and $X_i$ is

$$n_i={\rho X_i \over A_i m_H},$$

\noindent where $\rho$ is the density, $A_i$ the atomic weight
expressed in units of proton mass $m_H$. 

We are interested in resolving the diffusion equation in
spherically symmetric systems.
Let us consider a one-dimensional problem, where particles $i$ may diffuse along
the radial direction $r$. Thus

\begin{equation}
<V_i> =-{D \over X_i} {\partial X_i \over \partial r}.
\label{Vi}
\end{equation}

\noindent The variation of the number of particles in the element of volume $\upsilon$
and of surface $S$,
due only to diffusion, is

\begin{equation}
{\partial \over \partial t} \int_{\upsilon} n_i {\rm d}\upsilon =-\int_{S} n_i <\vec V_i>\cdot {\rm d}{\vec S}.
\label{nicont}
\end{equation}

\noindent From the divergence theorem and the expression of $<V_i>$ just seen above one obtains 

\begin{equation}
{\partial n_i \over \partial t}= -{1 \over r^2} {\partial \over \partial r} (r^2 n_i <V_i>) ={1 \over r^2} {\partial \over \partial r} 
\left (\rho r^2 D {\partial \left({n_i\over \rho}\right) \over \partial r}\right ).
\label{difni}
\end{equation}

\noindent Replacing $n_i$ by its expression as a function of $X_i$
in Eq.~(\ref{difni}) gives

\begin{equation}
{\partial \over \partial t}(\rho X_i )={1 \over r^2} {\partial \over \partial r}(\rho r^2  D {\partial X_i \over \partial r}).
\label{difXi}
\end{equation}

\noindent Thus Eq.~(\ref{difXi})
results from the expressions for the $<V_i>$ (Eq.~\ref{Vi}) and from the continuity equation
for the $n_i$'s (Eq.~\ref{nicont}).

Sometimes, Eq.~(\ref{difXi}) is written as below  
\begin{equation}
{\partial \over \partial t}(\rho c_i )={1 \over r^2} {\partial \over \partial r}(\rho r^2 D {\partial c_i \over \partial r}),
\label{difci}
\end{equation}

\noindent with partial concentration $c_i$ instead of mass fraction,
without indication on the precise meaning of what concentration means.
Partial concentration in
mass is equivalent to the mass fraction, while partial concentration in number is 
equal to $n_i/n$ where $n=\sum_i n_i$. 
Replacing $c_i$ 
in Eq.~(\ref{difci}), by $n_i/n$ and expressing $n_i$ as a function of $X_i$,
one obtains

\begin{equation}
{\partial \over \partial t}(\rho \mu X_i )={1 \over r^2} {\partial \over \partial r}\left (\rho r^2 D {\partial  \over \partial r}(\mu X_i)\right ),
\label{difnin}
\end{equation}

\noindent where $\mu=\rho/(n m_H)$ is the mean molecular weight of the ions.
This equation is identical to Eq.~(\ref{difXi}) only when $\mu$
remains constant as a function time and is constant as a function of $r$. Thus 
Eq.~(\ref{difnin}) is equivalent to Eq.~(\ref{difXi})
only when applied to minor constituents, the abundances of which
do not affect $\mu$ and when $\mu$ has no gradient. 
In more general cases, Eq.~(\ref{difnin}) is not equivalent
to Eqs.~(\ref{difni}) and (\ref{difXi}). 

This can also be seen in a slightly different way.
When, in Eq.~(\ref{difci}),
the $c_i$'s are identified with
the number fractions, one obtains

\begin{equation}
{\partial  \over \partial t}\left( \rho {n_i \over n}\right )
=-{1 \over r^2} {\partial \over \partial r} \left(r^2 \rho {n_i \over n} w_i   \right),
\label{difnin2}
\end{equation}

\noindent where we have introduced a velocity $w_i=-{D \over n_i/n} {\partial \over \partial r} \left ( {n_i \over n}\right )$.
This velocity is not a diffusive velocity in the sense that $\sum_i m_i n_i w_i=-nD m_H \partial \mu/\partial r$
is not equal to zero except in zones where $\mu$ is constant. 
Eq.~(\ref{difnin2}) expresses the change of $\rho n_i/n=\mu n_i m_H$ in an element of volume resulting from
the transport of the quantity $\rho n_i/n$. But physically, this is not $\rho n_i/n$ which diffuses
but the particles themselves. As above, there are two conditions for Eq.~(\ref{difnin2}) to be
equivalent to Eqs.~(\ref{difni}) and (\ref{difXi}): 1)
the particles which diffuse must
have a very small abundance, so small that their diffusion does not affect $\mu=\rho/(n m_H)$; 2) and $\mu$ does
not vary with the radius.
We conclude thus that when the diffusive transport is described by an
equation of the form given in Eq.~(\ref{difci}), the concentrations are equivalent to mass fractions.
The identification with number fractions results in unphysical description of the process
except in very particular situations.

Starting
from Eq.~(\ref{difXi}) and summing
over all the chemical species $i$, one obtains

$${\partial \rho \over \partial t}={1 \over r^2} {\partial \over \partial r}\left(\rho r^2  D {\partial  \over \partial r}(1) \right)=0.$$

\noindent This is quite consistent with the fact that the diffusive velocities must compensate
each other. Thus the density can be put outside from the time derivative in Eq.~(\ref{difXi}):




\begin{equation}
	\rho{\partial X_i \over \partial t}\bigg|_{m_r} = {1 \over r^2} {\partial \over \partial r}
\left (\rho r^2 D {\partial X_i \over \partial r}\right ),
\label{difx}
\end{equation}

\noindent where $m_r$ is the langrangian mass coordinate. This is the equation
we shall numerically resolve in this paper.
The conditions at the centre and the surface are:

\begin{equation}
	{\partial X_i \over \partial r }\bigg|_{m_r=0}={\partial X_i \over \partial r }\bigg|_{m_r=M}=0,
\label{cobx}
\end{equation}

\noindent where $M$ is the total mass of the star.

The equation expressing the diffusion of the angular momentum is (see e.g. 
Endal \& Sofia \cite{En78})

\begin{equation}
	\rho {\partial (r^2 \Omega) \over \partial t}\bigg|_{m_r} = {1 \over r^2} {\partial \over \partial r}
\left (\rho r^4 {D'} {\partial \Omega \over \partial r}\right ),
\label{difo}
\end{equation}

\noindent where $\Omega$ is the angular velocity and ${D'}$ the diffusion coefficient for the angular momentum. 
In our numerical experiments, the radii do not change with time\footnote {Let us note
that in more realistic stellar models, the radii of course vary as a function of time,
however time steps can be chosen  sufficiently small for considering them to be constant
during one time step. The same can be said for the diffusion coefficient and other
structure variables as for instance the density.}. 
Eq.~(\ref{difo}) then becomes 

\begin{equation}
	\rho r^2{\partial \Omega \over \partial t}\bigg|_{m_r} = {1 \over r^2} {\partial \over \partial r}
\left (\rho r^4 {D'} {\partial \Omega \over \partial r}\right )={\rm div}\left(\rho r^2 {D'} \vec \nabla (\Omega)\right ).
\label{difor}
\end{equation}

\noindent The conditions at the centre and at the surface are

\begin{equation}
	{\partial \Omega \over \partial r }\bigg|_{m_r=0}={\partial \Omega \over \partial r }\bigg|_{m_r=M}=0.
\label{cobo}
\end{equation}

Let us notice that in general the transport equation for the angular momentum may contain other terms
expressing the advection of angular momentum by meridional circulation and the effects of magnetic braking.
As these last two terms bring no specific problems with respect to diffusion, we do not consider them here.
However, the advection of angular momentum needs a particularly careful treatment
as well (cf. Meynet \& Maeder \cite{MMV}).

\section{The three tested methods}

\subsection{The implicit finite elements method}

The detailed description of the implicit finite elements method used to resolve
the diffusion equation for the chemical species 
is presented in Schatzman et al. (\cite{Sch81}) (see the appendix by Glowinsky \& Angrand). 
We present here the procedure for the case of the
angular momentum diffusion. The basic idea of this method is to decompose the unknown
function, here $\Omega(m_r,t)$, as a linear combination of well chosen independent functions. 
This decomposition can be performed in many different ways. We
adopt here the same decomposition as in Schatzman et al. (\cite{Sch81}). 
Of course, the conclusions concerning the ability of this method to provide physical solutions
will only refer to this particular choice.  

We decompose the star in $K$ shells, with the langrangian mass coordinate 
of the i$^{th}$ mesh point being $m_i$ ($m_i$ is the mass inside the sphere
of radius $r_i$). In the following all the quantities with an indice $i$ are evaluated
at the mesh point $i$.
The shells are numbered from 1 at the surface to $K$ at
the centre.
Let us introduce $K$ functions $b_i(m_r)$ defined by

$$
\begin{array}{clll}
         &            &{m_r-m_{i-1} \over m_i-m_{i-1}}  & {\rm if}\ \  m_r \in [m_{i},m_{i-1}],  \\
         &\nearrow    &                                  &                                        \\
 b_i(m_r)=&\rightarrow &{m_r-m_{i+1} \over m_i-m_{i+1}}  & {\rm if}\ \  m_r \in [m_{i+1},m_{i}],  \\
         & \searrow   &                                  &                                        \\
	 &            & 0                               & {\rm if}\ \  m_r \notin [m_{i+1},m_{i-1}]. \\
\end{array}
$$

\noindent The function $b_i$ is equal to one at $m_r=m_i$, is equal to zero at $m_{i+1}$ and $m_{i-1}$ and
varies linearly as a function of $m_r$ inbetween. 
By multiplying Eq.~(\ref{difor}) by each of the functions $b_i(m_r)$, one obtains
$K$ equations. The integration over the whole volume of the star, $V$, of each of these $K$
equations gives
\begin{equation}
	\int_V \rho r^2 {\partial \Omega \over \partial t} b_i {\rm d} V =
	\int_V {\rm div}\left(\rho r^2 {D'}{ \rm \bf grad}(\Omega) \right)b_i {\rm d}V.
        \label{feintv}
\end{equation}

\noindent Using the general relations: 
\begin{equation}
	{\rm div}(a{\bf v})=a {\rm div}({\bf v})+{ \rm \bf grad}(a)\cdot{\bf v},
\end{equation}
\begin{equation}
	\int_V {\rm div}({\bf v}){\rm d}V=\int_S {\bf v}\cdot{\rm d}{\bf S},
\end{equation}
\noindent where $a$ is a scalar, ${\bf v}$ a vector and $S$ the surface
of the volume $V$, one obtains that the right hand term of Eq.~(\ref{feintv}) becomes

$$
	\int_S \rho r^2 {D'} { \rm \bf grad}(\Omega) b_i\ {\rm d}{\bf S}-
	\int_V \rho r^2 {D'} {\rm \bf grad}(\Omega)\cdot {\rm\bf 
	grad}(b_i)\ {\rm d}V.
$$

\noindent The integral on the surface is null, since ${\rm \bf grad}(\Omega)={\bf 0}$
on the surface. Thus one has $K$ equations of the type

\begin{equation}
	\int_V \rho r^2 {\partial \Omega \over \partial t} b_i {\rm d}V+
	\int_V \rho r^2 {D'}{\partial \Omega \over \partial r}
	{\partial b_i \over \partial r} {\rm d}V=0.
\end{equation}

\noindent Using ${\rm d}m_r=\rho {\rm d}V=4\pi r^2 \rho {\rm d}r$, one obtains

\begin{equation}
	\int_0^M  \left[ r^2 {\partial \Omega \over \partial t} b_i +r^2
        {D'}(4\pi r^2 \rho)^2 {\partial \Omega \over \partial m_r}
	{\partial b_i \over \partial m_r}\right ] {\rm d}m_r=0. \label{feinte}
\end{equation}

\noindent The functions $b_i$ constitute a set of independent functions, therefore the unknown function $\Omega(m_r, t)$ can be expressed as a linear combination of $b_i$'s. One can write

\begin{equation}
	\Omega(m_r, t)=\sum_{j=1}^{K}\Omega_j b_j(m_r), \label{fedeco}
\end{equation}

\noindent where $\Omega_j=\Omega(m_j, t)$.
Equation~(\ref{fedeco}) simply says that to obtain $\Omega(m_r, t)$, with $m_r \in [m_{i+1},m_i]$, one simply interpolates linearly as a function of mass between
$\Omega_{i+1}$ and $\Omega_i$.

Expressing $\Omega(m_r, t)$ as indicated in Eq.~(\ref{fedeco}), Eq.~(\ref{feinte}) becomes

\begin{equation}
	\sum_j A_{ij}{\partial \Omega_j \over \partial t} +\sum_j B_{ij} \Omega_j =0.
        \label{finedis}
\end{equation}

\noindent where 
\begin{equation}
	A_{ij}=\int_0^M r^2 b_i b_j\  {\rm d}m_r,
\end{equation}
\begin{equation}
	B_{ij}=\int_0^M r^2 {D'} (4\pi r^2 \rho)^2 {\partial b_i \over \partial m_r}
	{\partial b_j \over \partial m_r}\ {\rm d}m_r.
\end{equation}

\noindent One has used the facts that the $\Omega_j$'s do not depend on $m_r$
and the $b_j$'s do not depend on time. When $|j-i| > 1$, there is no overlap
between the functions $b_i$ and $b_j$, thus
\begin{equation}
	A_{ij}=0,\ \ \ {\rm for}\ \ \ |j-i| > 1.
\end{equation}
\noindent For a given value of $i$ only, $A_{i,i-1}$, $A_{i,i}$ and
$A_{i+1,i}$ are different from zero. The same is true for the $B_{ij}$. 
Let us estimate $A_{i,i-1}$:
\begin{equation}
	A_{i,i-1}=\int_{m_{i}}^{m_{i-1}} r^2 {m_r-m_i \over m_{i-1}-m_i}{m_r-m_{i-1} \over m_{i}-m_{i-1}}\ {\rm d}m_r,
\end{equation}
\noindent one approximates $r^2$ by $0.5\left(r^2(m_{i-1})+r^2(m_i)\right)$, then, by trivial
integration, one obtains
\begin{equation}
	A_{i,i-1}=-0.5\left ( r^2(m_{i-1})+r^2(m_i)\right ){m_i-m_{i-1}\over 6}.
\end{equation}
\noindent The other matrix elements are obtained in the same way:
\begin{equation}
	A_{i+1,i}=-0.5\left ( r^2(m_{i+1})+r^2(m_i)\right ){m_{i+1}-m_i\over 6},
\end{equation}
\begin{equation}
	A_{i,i}=2(A_{i,i-1}+A_{i+1,i}).
\end{equation}

\noindent At the centre and at the surface, one has

$$
\begin{array}{ll}
	A_{K,K}  &=0.5 r^2(m_{K-1}){m_{K-1}\over 3},     \\
	A_{K,K-1}&=0.5 A_{K,K},                         \\
	A_{2,1}  &=0.5 A_{1,1},                         \\
	A_{1,1}  &=0.5\left ( r^2(1)+r^2(2)\right ){m_1-m_2 \over 3}.
\end{array}
$$

\noindent Let us set $\overline D=r^2 (4 \pi r^2 \rho)^2D'$ and $\overline D_{i,i-1}$ an appropriate
mean value of $\overline D$ between the shells $i$ and $i-1$
(see Sect.~4), then the $B_{i,j}$
becomes 
\begin{equation}
	B_{i,i-1}={\overline D_{i,i-1}\over m_i-m_{i-1}},
\end{equation}
\begin{equation}
	B_{i+1,i}={\overline D_{i+1,i}\over m_{i+1}-m_i},
\end{equation}
\begin{equation}
	B_{i,i}=-B_{i,i-1}-B_{i+1,i}.
\end{equation}

\noindent At the centre and at the surface, one obtains

$$
\begin{array}{ll}
	B_{K,K}  &={\overline D_{K,K-1}\over m_{K-1}},    \\
	B_{K,K-1}&=-B_{K,K},                             \\
	B_{2,1}  &=-B_{1,1},                             \\
	B_{1,1}  &={\overline D_{2,1}\over m_1-m_2}.
\end{array}
$$

\noindent Adopting an implicit discretisation in time, we have
for Eq.~(\ref{finedis})

\begin{equation}
	\sum_j A_{i,j}\left({\Omega_j^{n+1}-\Omega_j^n \over \Delta t} \right)
	+\sum_j B_{i,j}\Omega_j^{n+1}=0,
        \label{finedis2}
\end{equation}

\noindent or 
\begin{equation}
	\sum_j \left({A_{i,j}\over \Delta t}+B_{i,j} \right)\Omega_j^{n+1}=\sum_j {A_{i,j}\over \Delta t}\Omega_j^n,
\end{equation}

\noindent where $\Omega_j^n$ is equal to $\Omega(m_j,t^n)$. This linear system of equations is then solved by standard procedure.

\subsection{The Crank--Nicholson finite differences method}

Let us consider in a first step the case of the diffusion of the chemical elements.
Multiplying Eq.~(\ref{difx}) by $4\pi r^2$ and integrating with respect to
the spatial coordinate $r$ from
$r-\Delta r/2$ to $r+\Delta r/2$, one obtains

\begin{equation}
\Delta m {\partial X_i \over \partial t}\bigg|_{r}=\left( 4\pi \rho r^2 D {\partial X_i \over \partial r}\right ) \bigg|_{r-\Delta r/2}^{r+\Delta r/2},
\label{fdflu} 
\end{equation}

\noindent where $\Delta m=\int_{r-\Delta r/2}^{r+\Delta r/2} 4 \pi r^2 \rho dr$.
The quantity $\Delta r$ has been chosen sufficiently
small for  ${\partial X_i \over \partial t}$ to remain constant over the spatial range $\Delta r$ around $r$.
This integration reduces the equation to a first order differential equation, which simply expresses the fact that the time variation
of the mass of element $i$ in a spherical shell is equal to the difference between the mass
of element $i$ which enters in the shell and the mass of this same element which goes out
from the shell. As indicated in Sect. 2,
one can interpret the quantity,

\begin{equation}
-{D \over X_i} {\partial X_i \over \partial r}=-D {\partial \ln X_i \over \partial r}=<V_i>,
\label{vdifus}
\end{equation}

\noindent as a diffusion velocity, $<V_i>$, for the element $i$ (let us recall that the minus sign appears because,
when the spatial abundance gradient is positive, the velocity is directed inward).
In that case, Eq.~(\ref{fdflu}) writes
$$\Delta m {\partial X_i \over \partial t}\bigg|_r=-4\pi \rho r^2 X_i <V_i>|_{r-\Delta r/2}^{r+\Delta r/2}, $$
which makes clearly appear the difference of the mass fluxes between the two
borders of the shell. This is a more elementary description of diffusion.

One can also derive Eq.~(\ref{vdifus})
from the continuity equation. Indeed
supposing that the bulk gas velocity is zero, that $\partial \rho / \partial t =0$
(stationary situation), and that there is no sink/source terms of matter, the continuity equation becomes

\begin{equation}
\rho{\partial X_i \over \partial t}\bigg|_{r} =- {1 \over r^2} {\partial \over \partial r}( \rho r^2 <V_i> X_i).
\label{contin}
\end{equation}

\noindent By equating the right hand sides of Eqs.~(\ref{difx}) and (\ref{contin}),
one obtains for $<V_i>$ Eq.~(\ref{vdifus}) above.

\begin{figure}[tb]
  \resizebox{\hsize}{!}{\includegraphics[angle=0]{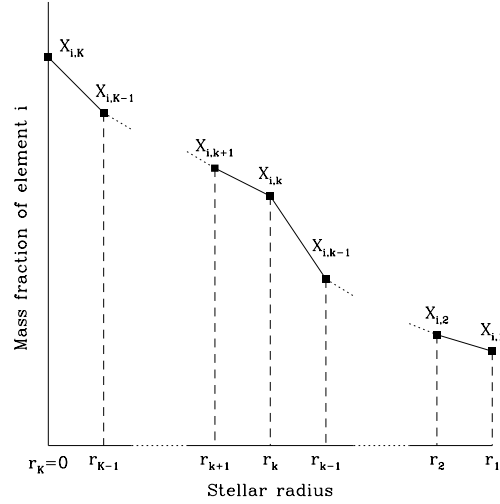}}
  \caption{Discretized distribution of the abundance of 
element $i$ in a star model as a function of radius. 
Only a few mesh points are represented. $K$ is the total number of mesh points.
Numbering increases from outside to inside.}
  \label{disc}
\end{figure}

Let us now consider the discretized abundance profile of element $i$ sketched in Fig. 1.
The mass of element $i$ removed from
point $k+1$ and added to point $k$ per unit time is (see Eq.~\ref{fdflu}):

\begin{equation}
-4 \pi r^2_{k+1/2} \rho_{k+1/2} D_{k+1/2} {\partial X_i \over \partial r}\bigg |_{k+1/2},
\end{equation}

\noindent where the physical quantities with a lower subscript $k+1/2$ are estimated at the midpoint between the mesh points $k+1$ and $k$. 
For instance, one takes $r_{k+1/2} =0.5(r_{k}+r_{k+1})$. For the diffusion coefficient, we use the
procedure exposed in Sect. 4.
A similar expression can be written for the mass of element $i$
removed from point $k$ and added to point $k-1$. Let us set $X_{i,k}$, the mass fraction of element $i$ evaluated at the mesh point $k$.
Thus one has,

\begin{eqnarray}
\lefteqn{\Delta m_k {\partial (X_{i,k}) \over \partial t}=} \nonumber \\
 & & - 4 \pi r^2_{k+1/2} \rho_{k+1/2} D_{k+1/2} {\partial  X_i \over \partial r}\bigg |_{k+1/2} \nonumber \\
 & &  +4 \pi r^2_{k-1/2}  \rho_{k-1/2} D_{k-1/2} {\partial X_i \over \partial r}\bigg |_{k-1/2},
\label{disspa}
\end{eqnarray}

\noindent where $\Delta m_k=\int_{r_{k+1/2}}^{r_{k-1/2}} 4 \pi r^2 \rho dr.$
Let us now discretize this equation as a function of time. The left hand term of Eq.~(\ref{disspa})
can be written

\begin{eqnarray}
\Delta m_k {\partial (X_{i,k}) \over \partial t}\rightarrow \Delta m_k {X_{i,k}^{n+1}-X_{i,k}^{n} \over \Delta t},
\end{eqnarray}

\noindent where the superscript $n$ indicates that the quantity is estimated at the time $t^n$.
The Crank--Nicholson 
method consists in evaluating the right hand term at the same time as the left hand term,
{\it i.e.} at time $t^{n+1/2}$. The system is  centered in time and thus second order accurate in time.
One obtains 

\begin{eqnarray}
\lefteqn{{X_{i,k}^{n+1}-X_{i,k}^{n} \over \Delta t}
=} \nonumber
\\ & &
-{\sigma_{k+1/2} \over \Delta m_{k}} D_{k+1/2} 
{{X_{i,k+1}^{n+1}+ X_{i,k+1}^{n}\over 2}-{ X_{i,k}^{n+1}+ X_{i,k}^{n}\over 2}\over r_{k+1}-r_{k}} \nonumber
\\ & &
+{\sigma_{k-1/2} \over \Delta m_{k}} D_{k-1/2} 
{{X_{i,k}^{n+1}+ X_{i,k}^{n}\over 2}-{ X_{i,k-1}^{n+1}+ X_{i,k-1}^{n}\over 2}\over r_{k}-r_{k-1}},
\label{distime}
\end{eqnarray}

\noindent where one has introduced the quantity $\sigma_{k+1/2}$ defined by

\begin{equation}
\sigma_{k+1/2}=4\pi r^2_{k+1/2}\rho_{k+1/2}.
\end{equation}

\noindent Let us define the quantity $[k+1,k]$ by

\begin{equation}
[k+1,k]={1 \over 2} \sigma_{k+1/2} {D_{k+1/2} \Delta t \over r_{k+1}-r_k},
\end{equation}

\noindent and let us separate the abundances evaluated at time $t^n$ from those estimated at time
$t^{n+1}$. Eq.~(\ref{distime}) then becomes

\begin{eqnarray}
{[k,k-1] \over \Delta m_k} X^{n+1}_{i,k-1}+
  \left (1- {[k,k-1] \over \Delta m_k}-{[k+1,k] \over \Delta m_k}\right )X_{i,k}^{n+1} & & \nonumber \\
  +{[k+1,k] \over \Delta m_k}X_{i,k+1}^{n+1}=\ \ \ \ \ \ \ \  & & \nonumber\\
-{[k,k-1] \over \Delta m_k} X^{n}_{i,k-1}+
  \left (1+ {[k,k-1] \over \Delta m_k}
+{[k+1,k] \over \Delta m_k}\right )X_{i,k}^{n} & & \nonumber\\
 -{[k+1,k] \over \Delta m_k}X_{i,k+1}^{n}\ .\ \ \ \ \ \ \ \  & &
\label{cni}
\end{eqnarray}

\noindent Following a similar line of reasoning, one obtains for the equations at the center and
at the surface (see Fig.~\ref{disc}),

\begin{eqnarray}
{[K,K-1] \over \Delta m_K }X_{i,K-1}^{n+1}+
\left (1-{[K,K-1] \over \Delta m_K }\right )X_{i,K}^{n+1}= & & \nonumber \\
 -{[K,K-1] \over \Delta m_K }X_{i,K-1}^{n}+
\left (1+{[K,K-1] \over \Delta m_K }\right )X_{i,K}^{n}\ , & &
\label{cnc}
\end{eqnarray}

\begin{eqnarray}
{[2,1] \over \Delta m_1 }X_{i,2}^{n+1}+
\left (1-{[2,1] \over \Delta m_1 }\right )X_{i,1}^{n+1}= & & \nonumber \\
 -{[2,1] \over \Delta m_1 }X_{i,2}^{n}+
\left (1+{[2,1] \over \Delta m_1 }\right )X_{i,1}^{n}\ , & &
\label{cnb}
\end{eqnarray}

\noindent where $\Delta m_K=\int_{0}^{r_{K-1/2}} 4 \pi r^2 \rho dr$ and
$\Delta m_1=\int_{r_{1-1/2}}^{r_1} 4 \pi r^2 \rho dr$.
Eqs.~(\ref{cni}), (\ref{cnc}) and (\ref{cnb}) constitute a system of linear equations whose unknowns are
the $X_{i,k}^{n+1}$. It is solved by using classical methods of 
tridiagonal matrix inversion.

The discretized equations describing the diffusion of the angular momentum
are obtained in a similar way. The final result are
equations similar to Eqs.~(\ref{cni}) (\ref{cnc}) and (\ref{cnb}) with $X_i$ replaced by $\Omega$,
$\Delta m_k$ replaced by $\Delta m_k r^2_k$, $\sigma_{k+1/2}$ replaced
by $\sigma_{k+1/2} r^2_{k+1/2}$ and $\sigma_{k-1/2}$ replaced
by $\sigma_{k-1/2} r^2_{k-1/2}$. Of course the diffusion coefficient must  be
the one describing the diffusion of the angular momentum.

Let us emphasize here that this method is not fully implicit
and thus, one can expect that some sort of Courant's condition
will limit its domain of validity (see Sect. 5).

\subsection{The implicit finite differences method}

In the implicit finite differences method,
the right hand term of Eq.~(\ref{disspa}) is estimated at time $t^{n+1}$ (in the explicit method
the right hand term would be estimated at time $t^n$). In this case, one obtains

\begin{eqnarray}
\lefteqn{{X_{i,k}^{n+1}-X_{i,k}^{n} \over \Delta t}
=} \nonumber \\
& & -{\sigma_{k+1/2} \over \Delta m_{k}} D_{k+1/2} 
{X_{i,k+1}^{n+1}- X_{i,k}^{n+1}\over r_{k+1}-r_{k}} \nonumber \\
& & +{\sigma_{k-1/2} \over \Delta m_{k}} D_{k-1/2} 
{X_{i,k}^{n+1}-X_{i,k-1}^{n+1}\over r_{k}-r_{k-1}},
\label{fully}
\end{eqnarray}

\noindent Let us define the bracket terms $[k+1,k]$ by

$$[k+1,k]=\sigma_{k+1/2} {D_{k+1/2} \Delta t \over r_{k+1}-r_k}\ ,$$

\noindent doing so, we obtain, separating the abundances at time $t^n$ from
those estimated at time $t^{n+1}$,

\begin{eqnarray}
\lefteqn{{[k,k-1] \over \Delta m_k} X^{n+1}_{i,k-1}+
  \left (1- {[k,k-1] \over \Delta m_k}-{[k+1,k] \over \Delta m_k}\right )X_{i,k}^{n+1}} \nonumber \\
& &  +{[k+1,k] \over \Delta m_k}X_{i,k+1}^{n+1}=
X_{i,k}^{n}\ .
\end{eqnarray}

\noindent Following a similar line of reasoning, one obtains for the equations at the center and
at the surface,

\begin{equation}
{[K,K-1] \over \Delta m_K }X_{i,K-1}^{n+1}+
\left (1-{[K,K-1] \over \Delta m_K }\right )X_{i,K}^{n+1}= X_{i,K}^{n}\ ,
\end{equation}

\begin{equation}
{[2,1] \over \Delta m_1 }X_{i,2}^{n+1}+
\left (1-{[2,1] \over \Delta m_1 }\right )X_{i,1}^{n+1}=X_{i,1}^{n}\ .
\end{equation}

One can easily check that the system of discretized equations
(both in the cases of the Crank--Nicholson scheme and
in the implicit finite differences method)
conserves the integrated
quantities of the elements over the mass of the star.
The system also keeps equal to one the sum of the $X_i$
and does not induce any diffusion 
when the chemical gradients are flat.

For obtaining the equations for the angular momentum, we apply the same recipe as indicated at the end
of the previous section. 
The system of equations describing the transport of the
angular momentum also conserves the total angular momentum.

\section{Estimate of $D_{k-1/2}$}

\begin{figure}[tb]
  \resizebox{\hsize}{!}{\includegraphics[angle=0]{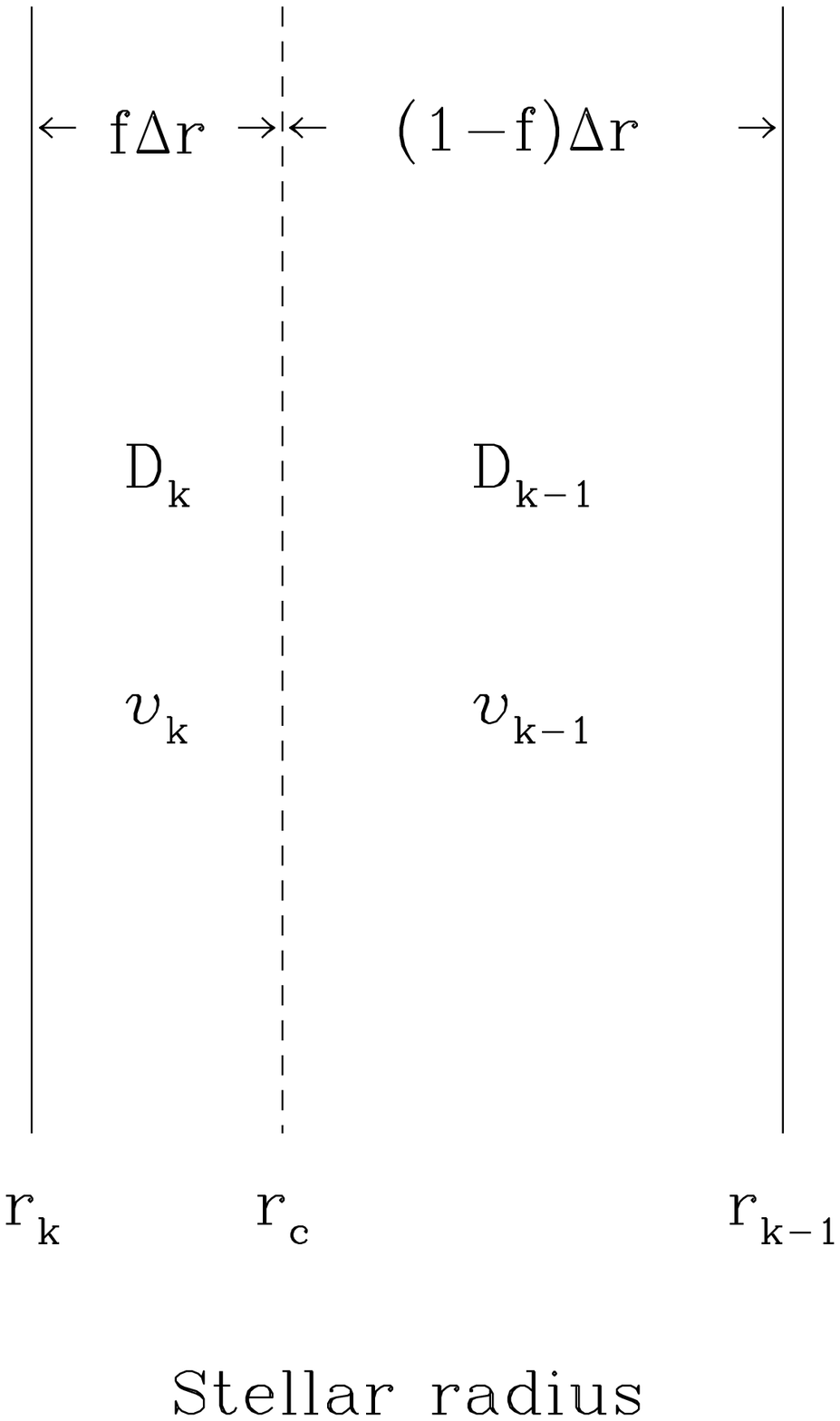}}
  \caption{Schematic representation of the mass shell between the radii $r_{k}$ and $r_{k-1}$. The
diffusive coefficient $D_{k}$ operates in the zone between
$r_{k}$ and $r_c$. The
diffusive coefficient $D_{k-1}$ operates in the zone between
$r_{c}$ and $r_{k-1}$. The quantity $\Delta r$ is equal to $r_{k-1}-r_k$.
}
  \label{schem}
\end{figure}

Let us consider two mesh points as in Fig.~\ref{schem}.  Let us suppose that the diffusion 
coefficient $D_{k-1}$ is the
coefficient for the whole region between $r_{k-1}$ and $r_{c}$. Similarly
the diffusion 
coefficient $D_{k}$ is the
coefficient for the whole region between $r_{c}$ and $r_{k}$.
When both mesh points are in a radiative or convective region,
the radius $r_c$ is equal to $(r_{k-1}+r_{k})/2$, otherwise $r_c$ is taken as
the radius of the limit between the radiative and the convective zone.

Over the path $f\Delta r$ (see Fig.~\ref{schem}), the particles have an average diffusive velocity $V_{k}$
(see Eq.~\ref{vdifus}) and over the path $(1-f)\Delta r$ they have 
a diffusive velocity $<V_{k-1}>$. The total time for going from $k$ to $k-1$
is equal to 

\begin{equation}
\Delta t = f\Delta r/<V_{k}> + (1-f)\Delta r/<V_{k-1}>. 
\end{equation}

\noindent The mean
velocity $<V_{k-1/2}>$ over the whole interval is given by $\Delta r/\Delta t$, thus one has
$$<V_{k-1/2}>={1 \over {f \over <V_{k}>} + {(1-f) \over <V_{k-1}>}}$$
$$={<V_{k}> <V_{k-1}>\over f <V_{k-1}>+(1-f)<V_{k}> }\ .$$

\noindent Now, the diffusive coefficient is proportional to the diffusive velocity
(see Eq.~\ref{vdifus}). This suggests the way to compute the diffusive coefficient between
two shells:

\begin{equation}
D_{k-1/2}={D_{k-1} D_{k}           
\over
f D_{k-1}  + (1-f) D_{k}}.
\label{meand}
\end{equation}

\noindent This expression is more physical than 

$$D_{k-1/2}=(D_{k-1}  + D_{k})/2,$$

\noindent  that
simply averages the diffusion coefficients. Equation~(\ref{meand}) implies 
that if $D_{k} >> D_{k-1}$ and $(1-f)$ is of the order of 0.5 then
$D_{k-1/2} \sim D_{k-1}$.
As expected, the smallest diffusion coefficient governs the diffusion between the two mesh points.
The simple algebraic mean would give $D_{k-1/2} \sim D_{k}/2$ a much greater diffusion coefficient,
which is not physically justified.
One sees also that
when $D_{k-1} = D_{k} = D$ then $D_{k-1/2}=D$ whatever the value
of $f$. In Sect.~5 below we shall illustrate by a numerical example
the importance of correctly evaluating the diffusion coefficient at the interface
between a convective and a radiative zone.

When the finite elements method is used, we need to evaluate the mean value between
two mass shells of the diffusion coefficient, $D$, multiplied by a factor $\alpha$,
which for the transport of the chemical elements is $(4\pi r^2 \rho)^2$ and for that of the angular 
momentum is $r^2(4\pi r^2 \rho)^2$ (see Sect. 3.1). In that case, one adopts the
following procedure:

$$
\alpha_{k-1/2} D_{k-1/2}=0.5\left (\alpha_{k}+\alpha_{k+1} \right){D_{k-1} D_{k}           
\over
f D_{k-1}  + (1-f) D_{k}}.
$$

\section{Tests and comparisons}

\subsection{Initial configuration and Courant's condition}

In this section, we study how the different methods described above behave in a very simple
configuration. Let us consider a uniform density sphere of 1 M$_\odot$, with
a radius of 1 R$_\odot$, composed of two chemical
elements $X$ and $Y$. The initial distribution of these two elements in the sphere is a quasi step function. 
The variation as a function of the radius of the abundance of one of these
two elements, let us call it $X$, at the beginning
of the computation is shown in Fig.~\ref{dfig} by the dotted line, 
(cf. also the dotted lines in Figs.~\ref{comp1} to \ref{comp4}). At the center, its initial value is 0.99,
in the outer regions $X=0.01$. 
The abundance $Y$ is simply $1-X$.
The diffusion coefficient is set equal to $10^{14}$ cm$^2$ s$^{-1}$
in the interior region ({\it i.e.} for a radius $r$ inferior to 0.7963 R$_\odot$ or a mass inferior to 0.5049 M$_\odot$) and to 1.5 10$^6$  cm$^2$ s$^{-1}$
in the outer zone. Such values for the diffusion coefficient imply that the interior has a very
small mixing timescale (of the order of $R^2/D$ $\sim$ 1 yr) and therefore will be always 
homogenized by mixing, while the outer one 
will have a much longer mixing timescale (of the order of 4.26 Myr). 
Such a situation is quite analog to a convective core
surrounded by a radiative envelope in a star. In the following we
shall define the interior region as ``the core'', and the outer one as ``the envelope''.
In our numerical experiments we keep the diffusion coefficient
constant as a function of time. The sphere is decomposed in 100 shells of
equal mass. 

From these data one expects that the whole star will be completely mixed
in less than about 10$^7$ yr and that the final abundance of element X
in the homogeneous star will be 0.5048.
From this initial structure, on can also estimate the ``Courant condition''.
Let us recall that the ``Courant condition'' imposes a superior limit $\Delta t_c$
to the time step for an explicit method to be stable (see e.g. Press et al. \cite{Teu92}). To estimate this limit,
one has to compute for each element and for each mass shell, 
the time required for the element to diffuse through 
the width of the shell {\it i.e.}

$$\Delta t_X = {\Delta r \over v_X} =  {\Delta r \over {D \over X}\left| {\partial X \over \partial r}\right|}.$$

\noindent Then one has to take the smallest value of all these
times. In a discretized form,  $\Delta t_c$ may be written

$$\Delta t_c = {\rm Min}\left ({(r_i-r_{i+1})^2 \over {D_{i+1/2}}\left| {X_{i}-X_{i+1} \over {(X_i+X_{i+1}) \over 2}}\right|}\right ).$$

For the simple initial structure considered here, the diffusion time
through a shell takes a non infinite value only at the interface between
the core and the envelope. In this case, one has that
$r_{i}-r_{i+1}$ is equal to 0.000199 R$_\odot$, $D_{i+1/2}$, estimated from Eq.~(\ref{meand}) with $f$ taken equal to 0.5, is equal to
3 10$^6$ cm$^2$ s, $X_{i+1}-X_{i}$ = 0.98 and ${X_i+X_{i+1} \over 2}$ is equal
to 0.5. This gives

$$\Delta t_c \sim 1\  {\rm yr.}$$

Let us stress that implicit methods are generally stable for
any time step, and thus are not limited by the Courant condition exposed above
(cf. Press et al.~\cite{Teu92}).
In the following we shall test this point.

\begin{table}
\caption{Maximum value over the star of the quantity $\Delta \chi=1-X(k)-Y(k)$ and of $\Delta B={B_{\rm final}-B_{\rm initial} \over 
B_{\rm initial}},$
for different values of $\tau$, $\Delta t$ (in years) and for different numerical schemes. $X$ and $Y$ are the mass fraction of the two elements composing the ``star'', $B$ is the
total angular momentum of the star. The labels CN and FI are for
``Crank--Nicholson'' and ``Fully Implicit'' respectively (see text).} \label{table}
\begin{center}\scriptsize
\begin{tabular}{|cc|cc|cc|cc|}
\hline
          &              &              &                  &                 &                 &                  &              \\
$\tau$    &  $\Delta t$  &\multicolumn{2}{|c|}{finite elem.}  & \multicolumn{2}{|c|}{finite diff.} & \multicolumn{2}{|c|}{finite diff.}\\
          &              &        &                  &   \multicolumn{2}{|c|}{CN}              & \multicolumn{2}{|c|}{FI}  \\
          &              &              &                 &               &                   &           &                     \\
          &  &  $\Delta \chi$ & $\Delta B$       &   $\Delta \chi $ & $\Delta B$        &  $\Delta \chi$  & $\Delta B$      \\
          &              &  [$10^{-9}$] &    [$10^{-4}$]   &      [$10^{-9}$]&     [$10^{-4}$]  &  [$10^{-9}$] &     [$10^{-4}$]                 \\
\hline
          &              &              &                  &                 &                  &              &                   \\
$10^4$    & 0.5          & 4.84         &  -0.99           &   0.40          &  -0.08           & -1.72        &   -0.08           \\      
$10^4$    & 1            & 4.84         &  -0.99           &   0.30          &  -0.08           & -2.63        &   -0.08           \\      
$10^4$    & 10           & 4.84         &  -0.99           &  -0.85          &  -0.08           & -0.66        &   -0.08           \\      
$10^4$    & 10$^2$       & 4.84         &  -0.99           &   0.86          &  -0.08           & -2.14        &   -0.08           \\      
$10^5$    & 10           & 1.85         &  -1.32           &  -7.92          &  -0.23           & -6.24        &   -0.23           \\      
$10^5$    & 10$^2$       & 1.85         &  -1.32           &  7.68           &  -0.23           &  -20         &   -0.23           \\      
$10^5$    & 10$^3$       & 1.85         &  -1.32           &  8.76           &  -0.23           & -8.05        &   -0.23           \\      
$10^6$    & 10$^2$       & -81          &  -2.01           &  64             &  -0.59           &  -168        &   -0.59           \\      
$10^6$    & 10$^3$       & -81          &  -2.01           & 69              &  -0.59           &  -67         &   -0.59           \\      
$10^7$    & 10$^3$       & -498         &  -2.32           & 497             &  -0.79           &   -484       &   -0.80           \\      
$10^7$    & 10$^4$       & -498         &  -2.33           &  397            &  -0.78           & 1340         &   -0.79           \\      
$10^8$    & 10$^4$       & 29542        &   2.24           &  3772           &  -0.67           & 12426        &   -0.80           \\      
$10^9$    & 10$^4$       & 245311       &  -1.34           &  37871          &   0.44           & 123274       &   -0.90           \\      
$10^9$    & 10$^5$       & 245311       &  -1.72           &  25155          &  -0.65           & 48542        &   -0.80           \\      
          &              &              &                  &                 &                  &              &                   \\
\hline
\end{tabular}
\end{center}

\end{table}

\subsection{Comparisons of the methods}

\begin{figure}[tb]
  \resizebox{\hsize}{!}{\includegraphics[angle=0]{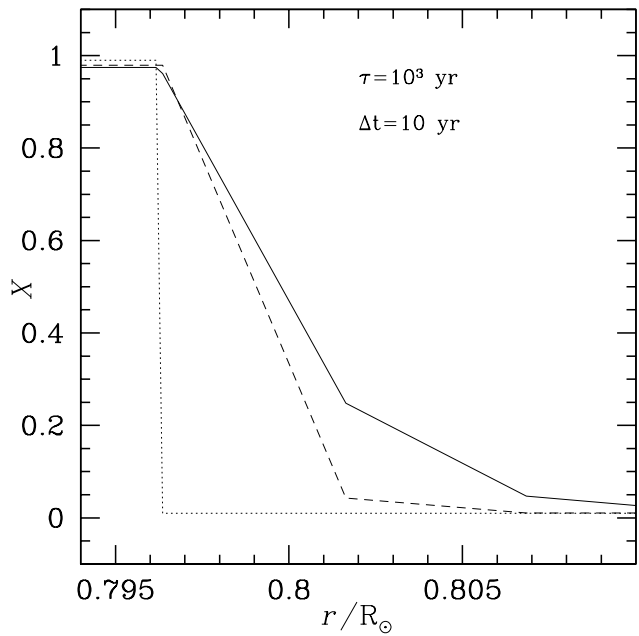}}
  \caption{Comparison of the abundances obtained with the implicit finite differences method
using two different ways of evaluating the diffusion coefficient $D_{k-1/2}$
between two
mesh points (see Sect. 4). The dotted line shows the 
initial situation ($\tau$=0). A time step $\Delta t$ of 10 years was adopted and
the computation was stopped at an age of $\tau =$ 1000 years.
The dashed line show the result obtained when
one takes for $D_{k-1/2}$ a simple algebraic mean, the continuous line
presents the result obtained using Eq.~(\ref{meand}).}
  \label{dfig}
\end{figure}

Let us begin by illustrating the importance of correctly estimating the diffusion coefficient at the
interface between a convective and a radiative zone.
On Fig.~\ref{dfig}, the resulting distributions of the element $X$ inside the star
after 1000 yr is shown. The time step used is $\Delta t$ = 10 yr
The continuous line shows the results obtained using Eq.~(\ref{meand}) for
$D_{k-1/2}$. The dashed line represent the solution obtained using 
a simple algebraic mean.  We see that the results
are significantly different, in particular the algebraic mean tends to slightly increase
the convective core. This may have important consequences when such increases
are repeatedly applied over the whole evolution of a star. The use
of Eq.~(\ref{meand}) which results from physical considerations is thus
recommended, and this the one we used in the numerical experiments we now describe.

\begin{figure}[tb]
  \resizebox{\hsize}{!}{\includegraphics[angle=0]{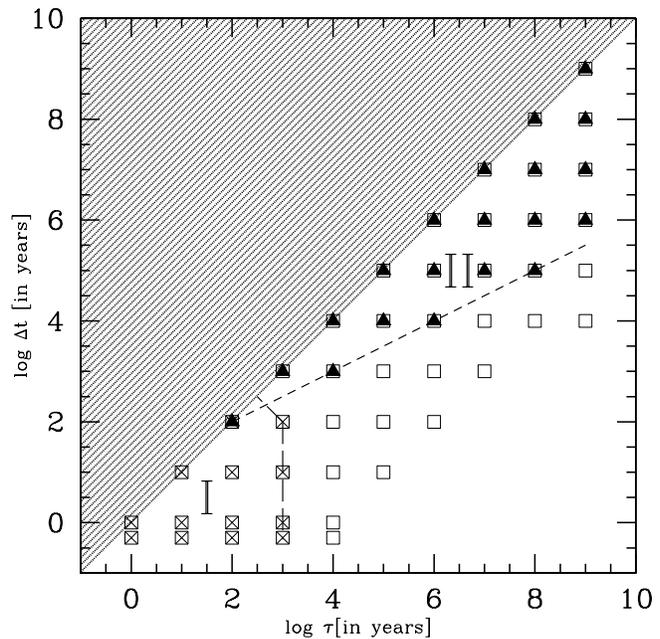}}
  \caption{Computations have been performed with the three methods
for each couple of ($\log \Delta t$, $\log \tau$) values corresponding
to the position of a square in the above figure; $\Delta t$ is the time step,
$\tau$ is the duration over which the computation was performed, (no computation, of course,
have been performed in the grey zone where $\log \Delta t > \log \tau$). 
The zone I (square with a cross inside) shows the ``non--physical'' region
for the implicit finite elements method. By ``non--physical'', we mean here that
negative abundances are obtained.
The zone II 
(square with a filled triangle inside) corresponds to the zone of ``non--physical'' solutions
for the  Crank--Nicholson finite differences method.
The implicit finite differences method gives physical solution in all the cases
considered here (indicated by squares filled or empty).
}
  \label{plan}
\end{figure}

We have computed the diffusion of the chemical elements
(and of the angular momentum, see Sect. 5.7) for
different durations  $\tau$ of the period during which the diffusion operates and for different time steps $\Delta t$.
In Fig.~\ref{plan} the set of values ($\log \Delta t$, $\log \tau$) explored are indicated. For each couple of values, we
performed computations with the three methods described above, namely the implicit finite elements method, the  Crank--Nicholson finite
differences method and the implicit finite
differences method. Most of the results are displayed in Figs.~\ref{comp1} to \ref{comp4}. On these figures, 
the results obtained for increasing durations, using the same time step,
are ordered horizontally (from left to right), while the results obtained for the same duration, with increasing time steps,
are arranged vertically (from top to bottom). Since the Courant time step is equal to one year, the time
step $\Delta t$ expressed in years gives directly the time step in units of the Courant time step.

\subsection{The implicit finite elements method}

The solutions obtained with the implicit finite elements method are shown in Figs.~\ref{comp1} and \ref{comp2}
(the dashed--dotted lines).
This method gives reasonable results in all the cases explored in this work except for small values
of $\tau$. 
We note that for $\tau < 1000$ yr, whatever is the time step, the solution leads to
negative values of $X$ just above the ``convective core''. 
If the integration is
performed over a sufficiently long time, the system eventually reaches a reasonable solution,
even when instabilities appear at earlier time
(see for instance in Figs.~\ref{comp1} and \ref{comp2} the evolution when $\tau$ increases with $\Delta t$ = 0.5 yr).
One notes that the star is completely mixed for $\tau > 10^7$, {\it i.e.} for durations
more than twice the mixing timescale of the envelope (about 4.3 Myr, see Sect.~5.1), which
is not very satisfactory. The final abundance
in the homogeneous star is on the other hand equal to that expected ($\sim 0.5$).

As expected from an implicit scheme, reasonable solutions are obtained even if the time steps
are much greater than 
the time step given by the Courant condition. 
Inspection of the results for $\tau \geq 1000$ yr show in general a great stability of the solution with respect to
the choice of the time step. 
The solutions obtained with the implicit
finite elements method are in general less mixed that those obtained with the implicit finite differences method
(see Figs.~\ref{comp3} and \ref{comp4}). 

\subsection{The Crank--Nicholson finite differences method}

This method is a kind of compromise between a fully implicit and an explicit scheme.
In order to better understand what happens here, let us briefly recall 
a few generalities about implicit and explicit schemes (see Press et al \cite{Teu92}, p. 838--842).

Explicit finite differences schemes are stable only if the
time steps satisfy the Courant condition. Such methods are not
suited for the computation of the secular evolution of stars. Indeed, we are interested
in modeling the evolution of features with spatial scales of the order of the radius of the star $R$, much greater
than the distance between two mesh points $\Delta r$. If we are limited to time steps satisfying
the Courant condition, we will need of the order of $R^2/(\Delta r)^2$ steps before
anything interesting begins to happen. We would like instead use timesteps of the order
of $R^2/D$ or, maybe, for purpose of accuracy, somewhat smaller. 

\begin{figure*}[tb]
  \resizebox{\hsize}{!}{\includegraphics[angle=0]{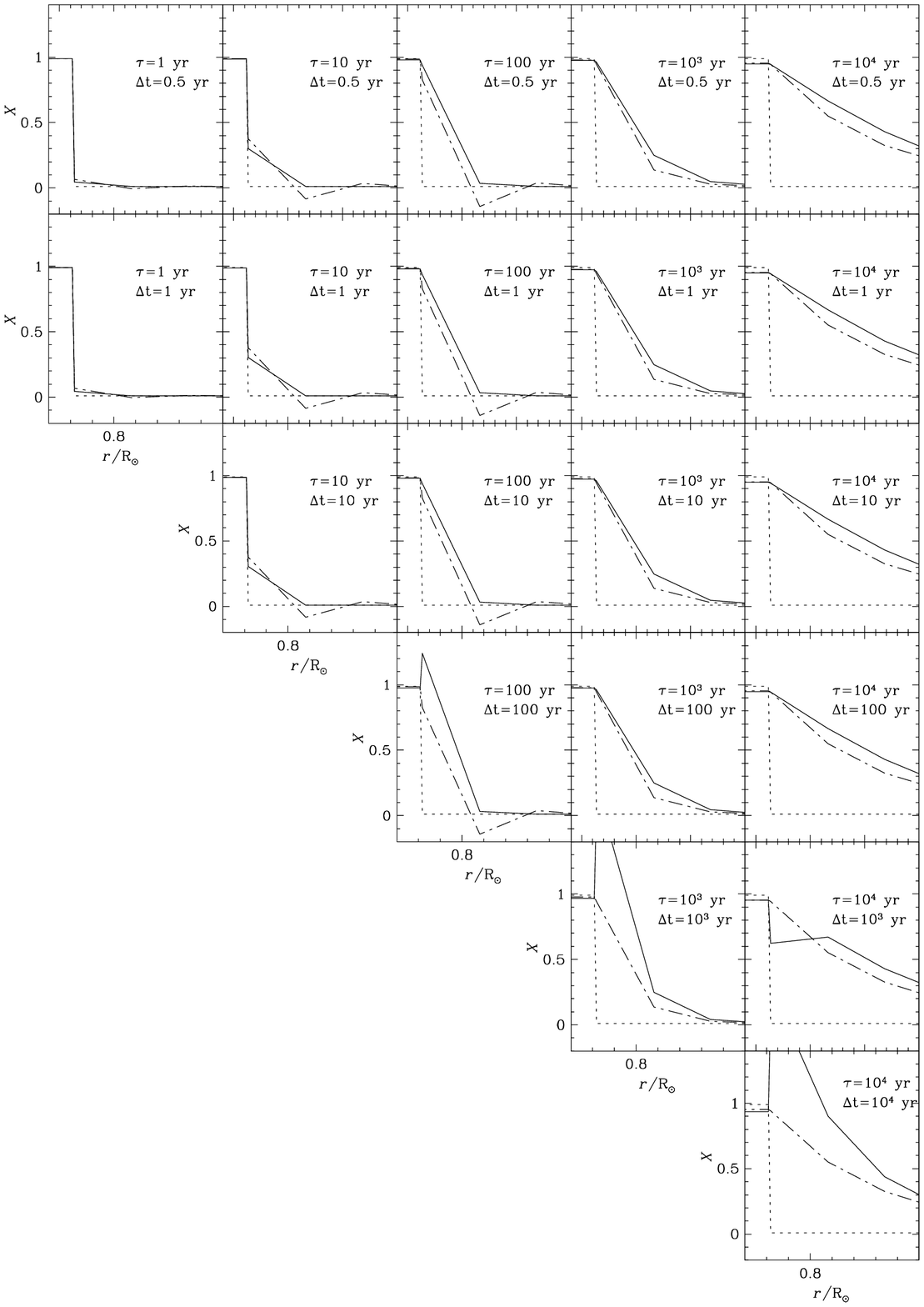}}
  \caption{Abundance of the element $X$ as a function of the distance to the center.
The dotted line show the situation at the beginning of the computation. The continuous line
refers to the solution obtained after a time $\tau$ by the Crank--Nicholson finite
differences method. The dashed--dotted line shows the
result obtained after a time $\tau$ by the implicit finite elements method. The time step $\Delta t$
used in each case is indicated. In all the cases, the Courant condition time is about 1 yr.
This means also that the value of $\Delta t$ is equal to the ratio $\Delta t/\Delta t_c$.
}
  \label{comp1}
\end{figure*}

\begin{figure*}[tb]
  \resizebox{\hsize}{!}{\includegraphics[angle=0]{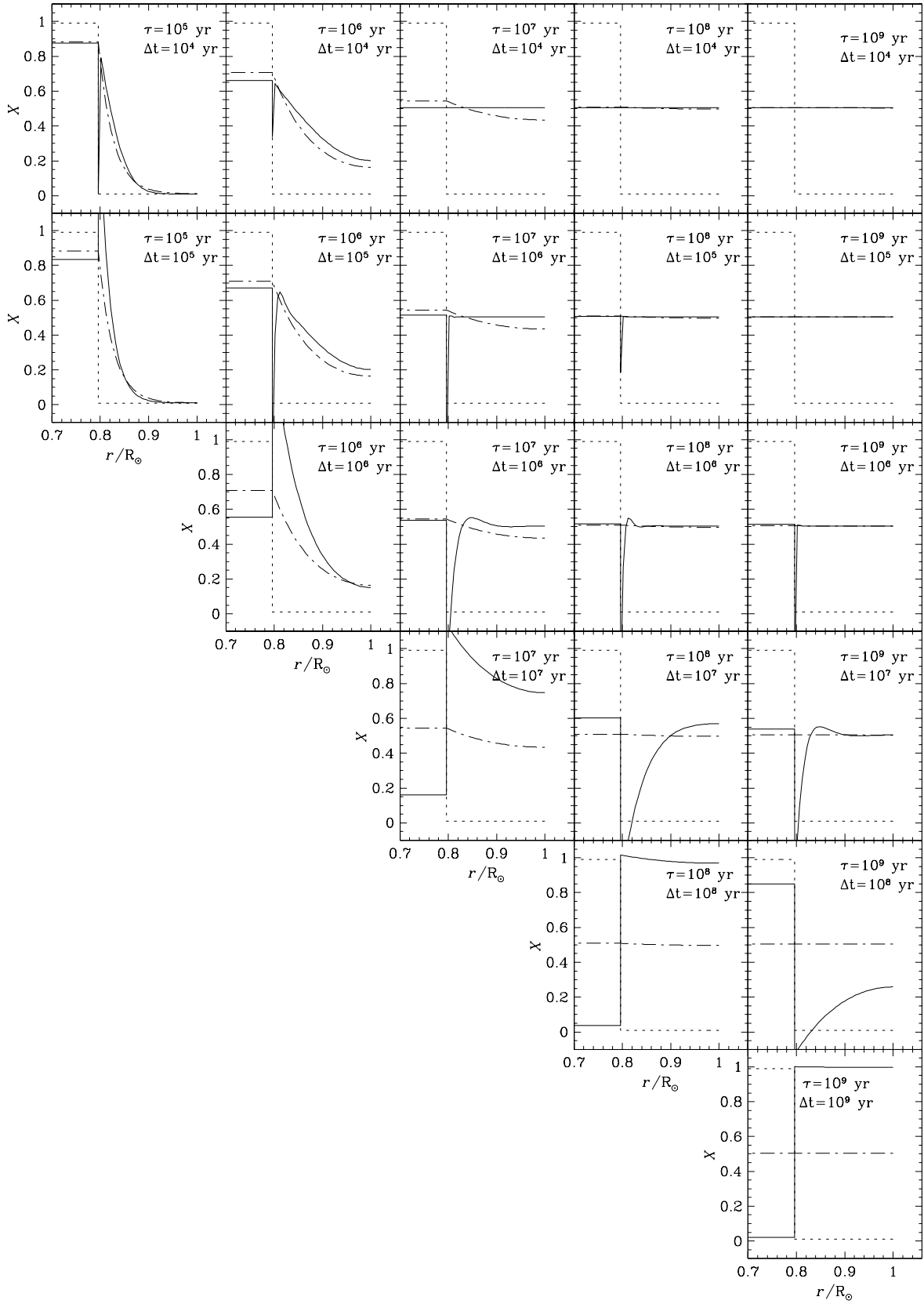}}
  \caption{Same as Fig.~5 for other values of $\tau$ and $\Delta t$.}
  \label{comp2}
\end{figure*}

With such great timesteps, it is no long possible to describe accurately what happens
at small spatial scales. However
at small scale, the differencing scheme must do ``something stable,
innocuous, and perhaps not too physically unreasonable'' write Press et al. (\cite{Teu92}).
One possibility is to use a Crank--Nicholson differencing scheme. One of its main property is to let small--spatial--scale features
maintain their initial amplitudes. In that case, according to Press et al. (\cite{Teu92}),
the evolution of the larger--scale features, in which we are interested in,
take place superposed with a kind of ``frozen in'' (though fluctuating)
background of small--scale features.
This is what happens in our numerical experiments. Indeed
looking at the continuous lines in Figs.~\ref{comp1} and \ref{comp2}, one sees that the amplitude of the initial 
step in chemical composition is more or less maintained, although with great fluctuations.

The method gives reasonable solutions for not too big time steps. More precisely,
physical solutions are obtained when $\Delta t$ in years  is inferior to $10\sqrt{\tau}$, or when $\tau$
in years is inferior to 100 yr
(see Fig.~\ref{plan}). This restriction on the time step is reminiscent of a kind
of Courant's condition. This is not so surprising given that the Crank--Nicholson scheme
is a mixture of both an implicit (not submitted to Courant condition) and 
an explicit method (submitted to Courant condition).

Let us finally note that in its domain of validity in the log $\Delta t$ versus log $\tau$ plane
this method gives the same solution as the implicit finite differences method, and
since, it is centered in time,
this scheme is second--order accurate in time.

\subsection{The implicit finite differences method}

Another possibility for imposing an ``inoccuous'' behaviour to the small--scale features
is to adopt an implicit method. Such schemes drive small--scale features to their equilibrium form,
{\it i.e.} imposes that at small scales

$$
{\partial \over \partial r}
\left (\rho r^2 D {\partial X_i \over \partial r}\right )\rightarrow 0.
$$

\noindent This can be seen from Eq.~(\ref{fully}) with $\Delta t \rightarrow \infty$.

\begin{figure*}[tb]
  \resizebox{\hsize}{!}{\includegraphics[angle=0]{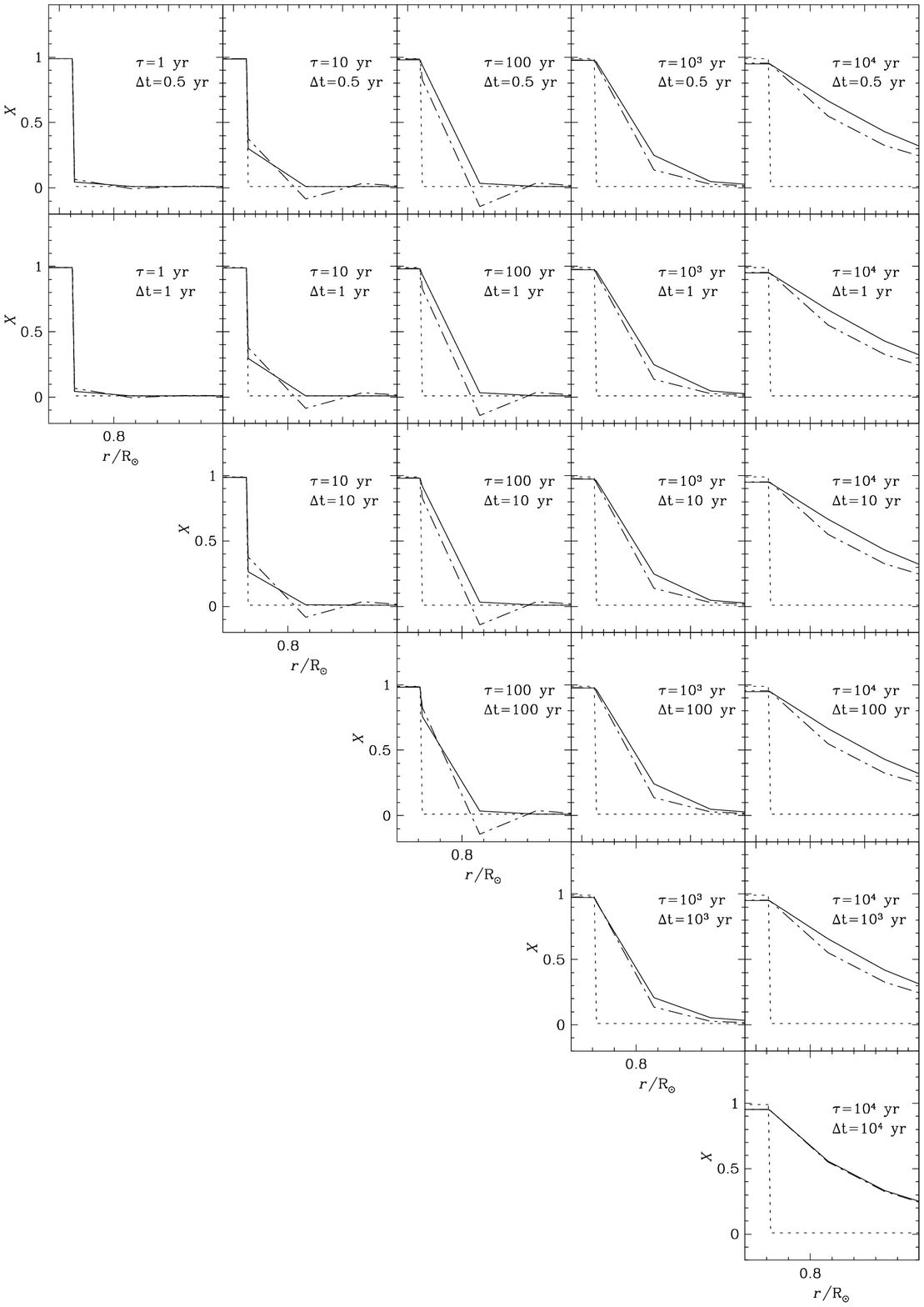}}
  \caption{Same as Fig.~5 except that the continuous line corresponds to the result
obtained after a time $\tau$ by the implicit finite
differences method.}
  \label{comp3}
\end{figure*}

\begin{figure*}[tb]
  \resizebox{\hsize}{!}{\includegraphics[angle=0]{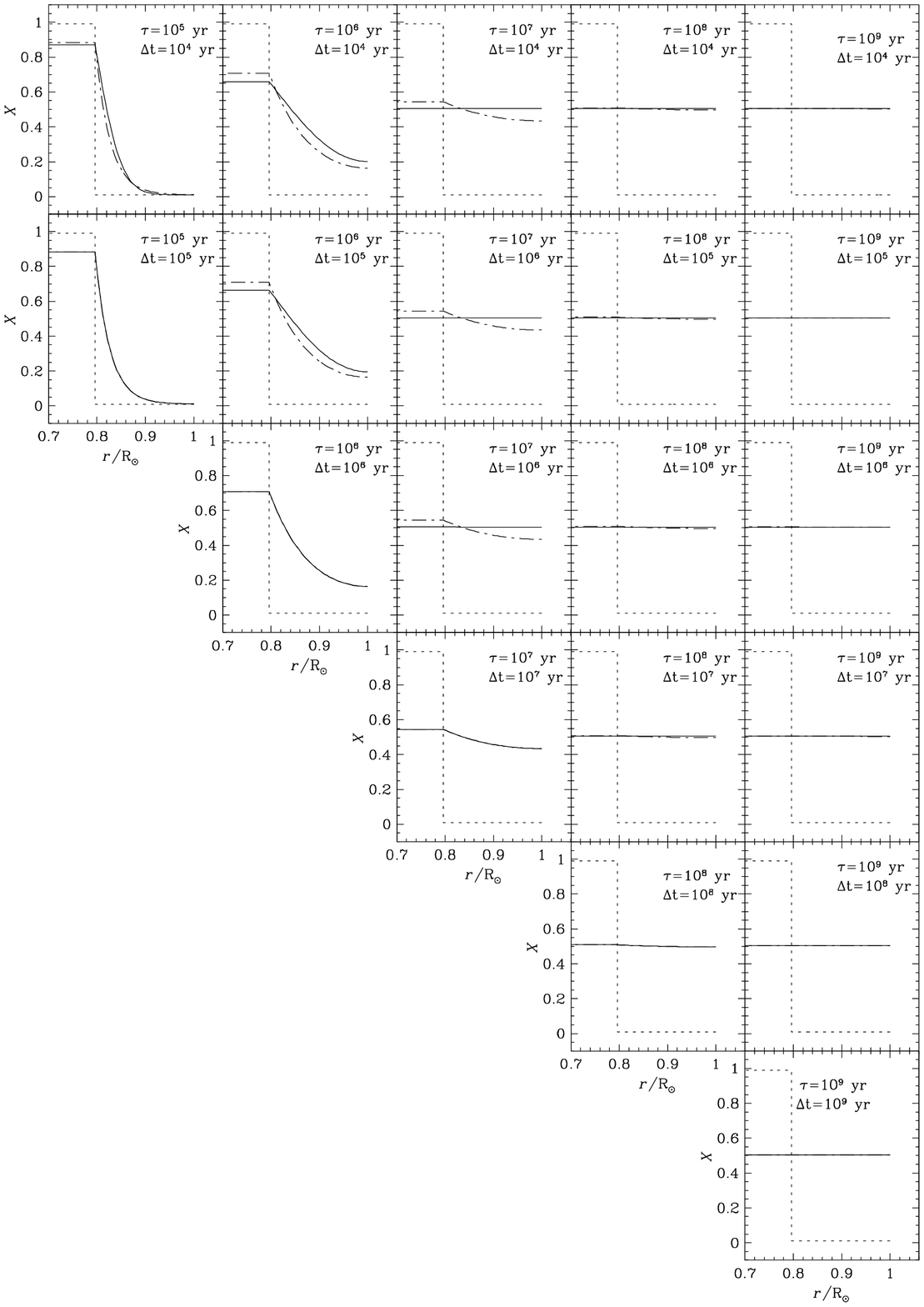}}
  \caption{Same as Fig.~6 except that the continuous line corresponds to the result
obtained after a time $\tau$ by the implicit finite
differences method. }
  \label{comp4}
\end{figure*}

The solutions obtained with the implicit finite differences method are shown in Figs.~\ref{comp3} and \ref{comp4}
(continuous lines). The results show in general a great stability of the solution with respect to
the choice of the time step.   
Only when the time step is of the same order of magnitude as $\tau$ can
we notice some differences (compare for instance the continuous line for $\tau$ = 10$^7$ yr and $\Delta t$ = 10$^6$ yr with
the case $\tau$ = 10$^7$ yr and $\Delta t$ = 10$^7$ yr in Fig.~\ref{comp4}).

This method leads to a completely mixed star after a time less than 10$^7$ in agreement
with the estimate made in Sect.~5.1. The final abundance of the element $X$ in the homogeneous
star is also equal to that expected.

Although, in that case, the differencing scheme is only first--order accurate in time,
it has the advantage over the Crank--Nicholson finite differences method and the implicit finite elements method
to propose a physical solution ({\it i.e.} without negative abundance values) for all the cases investigated here
(see Fig.~\ref{plan}). Moreover, as seen just above, it predicts a timescale for the star to be completely homogenized in agreement with
the usual analytical estimate (see Sect. 5.1).
In that respect this
method does appear as the best one.

\subsection{Conservation of the sum of the abundances}

Let us now have a look on the ability of the different schemes to conserve the sum of the
abundances (in mass fraction). In each shell and at any time step, one should have that $X$+$Y$=1.
In Table~\ref{table}, we indicate for all the cases where the three methods give physical solutions,
the maximum over the star of the quantity
$\Delta \chi=1-X(k)-Y(k)$ at the end of the computation. As expected  
we see that in general $\Delta \chi$ becomes greater when $\tau$ increases.
The Crank--Nicholson finite differences method appears to give, in most of the
cases, the smallest
values for $\Delta \chi$ (inferior to 10$^{-4}$ for the cases considered). This is likely related to the fact that this method is
second order accurate in time.
The implicit finite elements method gives in general the highest
values (in the worst case $\Delta \chi$ is of the order of 10$^{-3}$), while the implicit finite differences method gives in general results inbetween.
Thus
the implicit finite differences method enables to avoid unphysical solutions and keeps reasonably well the sum
of the abundances equal to one.

\subsection{Conservation of the angular momentum}

We have performed similar tests and comparisons for the diffusion of the angular momentum.
We started from an initial configuration where the ``convective core'' defined in Sect 5.1
has an angular velocity equal to 1$\cdot$10$^{-5}$ sec$^{-1}$ and the radiative envelope an
angular velocity equal to 1$\cdot$10$^{-6}$ sec$^{-1}$. The other variables were taken
as described in Sect.~5.1. In such situation the Courant time step, which writes

$$\Delta t_c = {\rm Min}\left( {(r_i-r_{i+1})^2 \over {D_{i+1/2}}\left| {\Omega_{i}-\Omega_{i+1} \over {(\Omega_i+\Omega_{i+1}) \over 2}}\right|}\right ),$$

\noindent is equal to 1.24 year, not very different from the one we obtained for the
diffusion of the chemical species. It is thus not surprising that 
the results we obtained are qualitatively similar to those presented in Figs.~\ref{comp1} to \ref{comp4}. 
The zones where unphysical solution are encountered 
are the same
as those presented in Fig.~\ref{plan}. We show in Table~\ref{table} the value of the quantity $\Delta B$ which
is equal to

$$\Delta B={B_{\rm final}-B_{\rm initial} \over 
B_{\rm initial}},$$

\noindent where $B_{\rm initial}$, $B_{\rm final}$  are the total angular momentum of the star at the beginning,
respectively at the end of the computation. 
The relative error on the total angular momentum is much greater than
the error on the sum of the abundances. This arises because the total angular momentum
is an integrated value over the whole system, implying that the errors can accumulate.
The sum of the abundances, instead, is evaluated locally in a given shell.

We see that the finite
differences method appear to better conserve the total angular momentum.
The Crank--Nicholson finite differences method gives the best results, followed by the implicit finite differences method which reaches the same
level of accuracy in most cases.
Thus the implicit finite differences method that we recommended on the base of the results obtained for the diffusion
of the chemical species gives also very satisfactory results for the diffusion of the angular momentum.

\section{Conclusion}

From the above numerical experiments, it appears that the implicit finite differences method
seems to be the most robust one, giving physical solutions in all the cases studied here spanning more than nine orders
of magnitude in $\tau$ and $\Delta t$. It has moreover 
the following characteristics: 1) it reduces the problem to a first order 
differential equation, 2) it enables an easy and clear interpretation of what happens 
physically in the system, 3) it is quite easy to implement in a code, 4) it conserves reasonably well the
sum of the abundances at each mesh point as well as the total angular momentum.

On the basis of the new tests and analysis made here, we can recommend this method to
resolve the diffusion equation in stellar interiors.
Of course many more
tests could be performed by changing the initial conditions. We restrained our
discussion here on a case with a very sharp gradient in order to test the different
numerical methods in some extreme conditions. Adopting an initially shallower gradients
will tend to make the things much more easier for all three methods and they
would give identical results. 

The diffusion coefficient between two mesh points has to be evaluated correctly
in order to obtain reliable results. The mean diffusion coefficient is equal
neither to the algebraic nor to the geometric mean. Its expression is given
in Eq.~(\ref{meand}).


\begin{thebibliography}{}
\bibitem[1996]{ba96} Battaner E. 1996, Astrophysical Fluid Dynamics, Cambridge
University Press
\bibitem[1992]{Cha92} Chaboyer B., Zahn J.--P. 1992, A\&A, 253, 173
\bibitem[1995]{Cha95} Charbonnel C. 1995, ApJ Letters, 453, 41
\bibitem[1999]{Deni99} Denissenkov P.A., Ivanova N.S., Weiss A., 1999, A\&A, 341, 181
\bibitem[1978]{En78} Endal, A.S., Sofia, S. 1978, ApJ, 220, 279
\bibitem[2000]{He20} Heger A., Langer N., Woosley S.E., 2000, ApJ, 528, 368
\bibitem[1995]{Ka95} Kawaler S.D. 1995, in Stellar Remnant, Saas Fee Advanced Course 25, Eds. G. Meynet
\& D. Schaerer, Springer, p. 1
\bibitem[2001]{MM01} Maeder A., Meynet G. 2001, A\&A, 373, 555
\bibitem[2000]{MMV}   Meynet G., Maeder A. 2000, A\&A, 361, 101
\bibitem[2003]{MM00} Meynet G., Maeder A. 2003, A\&A, 404, 975
\bibitem[1992]{Teu92} Press W.H., Teukolsky S.A., Vetterling W.T., Flannery B.P. 1992,
Numerical Recipes in Fortran, The Art of Scientific Computing, Cambridge
University Press
\bibitem[1981]{Sch81} Schatzman E., Maeder A., Angrand F., Glowinski R. 1981,
A\&A, 96, 1
\bibitem[1997]{Ta97} Talon S., Zahn J.P., 1997, A\&A, 317, 749
\bibitem[1998]{Vau98} Vauclair S., Charbonnel C. 1998, ApJ, 502, 372
\bibitem[1992]{Za92} Zahn J.--P. 1992, A\&A, 265, 115
\bibitem[2000]{zi00} Zienkiewicz O.C., Taylor R.L. 2000, The Finite Element Method,
Butterworth--Heinemann, Oxford
\end{thebibliography}
\end{document}